\documentclass[modern]{aastex61}
\usepackage{graphicx}% Including figure files
\usepackage{amsmath}% Advanced maths commands
\usepackage{amssymb}% Extra maths symbols
\usepackage{enumitem}

\newcommand\mybar{\kern1pt\rule[-\dp\strutbox]{.8pt}{\baselineskip}\kern1pt}

\setlist[itemize]{noitemsep, topsep=0pt, leftmargin=*}

\shorttitle{Deriving a Minimum Mass Density for a UFO}
\shortauthors{Loeb}

\begin{document}

\title{A New Method to Derive an Empirical Lower Limit on the Mass
  Density of a UFO}

\author{Abraham Loeb}

\affiliation{Head of the Galileo Project,
  Astronomy Department, Harvard University, 60 Garden St., Cambridge,
  MA 02138, USA}

\begin{abstract}
I derive a lower limit on the mass of an Unidentified Flying Object
(UFO) based on measurements of its speed and acceleration, as well as
the infrared luminosity of the airglow around it. If the object's
radial velocity can be neglected, the mass limit is independent of
distance. Measuring the distance and angular size of the object allows
to infer its minimum mass density. The {\it Galileo Project} will be
collecting the necessary data on millions of objects in the sky over
the coming year.

\end{abstract}

\section{Introduction}

Any object moving through air radiates excess heat in the form of
infrared airglow luminosity, $L$. The airglow luminosity is a fraction
of the total power dissipated by the object's speed, $v$, times the
frictional force of air acting on the object. The radiative efficiency
depends on the specific shape of the object and the turbulence and
thermodynamic conditions in the atmosphere around it~\citep{frict}. If the
object accelerates, then this friction force must be smaller than the
force provided by the engine which propels the object. The net
force equals the object's mass, $M$, times its acceleration, $a$.

In conclusion, one gets an unavoidable lower limit on the mass of an
accelerating object. The object's mass must be larger than the
infrared luminosity from heated air around it, divided by the product
of the object's acceleration and speed. In other words:
\begin{equation}
M > {L\over \vert v\times a \vert}~~.
\label{eq:limit}
\end{equation}
This limit provides an elegant way to constrain the minimum mass of
Unidentified Flying Objects (UFOs), also labeled as Unidentified
Anomalous Phenomena (UAPs; see \citet{Watters} for an overview). To
turn the inequality into an equality, one needs to know the detailed
object shape and atmospheric conditions around the
object.

The first {\it Galileo Project} Observatory at Harvard
University~\citep{Loeb} collects data on $\sim 10^5$ objects in the sky
every month. A comprehensive description of its commissioning data on
$\sim 5\times 10^5$ objects was provided in a recent
paper~\citep{Domine}. The data includes infrared images captured by an
all-sky {\it Dalek} array of eight uncooled infrared cameras placed on
half a sphere~\citep{Loeb,Watters}.

\section{Method}

Within the coming month, the {\it Galileo Project}’s research team plans to
employ multiple {\it Dalek}s separated by a few miles, in order to measure
distances to objects through the method of triangulation.

The heated air's infrared flux, $f$, and distance, $R$, can
be combined to infer the luminosity, $L$, through the relation:
\begin{equation}
  L= 4\pi f R^2~~.
\label{eq:lum}
\end{equation}  
 
The angular velocity, $(d\theta/dt)$, times the distance, $R$, provides
the transverse component of the full velocity vector, ${\bf v}$, which can
be combined with the time derivative of the distance, $(dR/dt)$, to get
the total speed,
\begin{equation}
  v=\left\{\left[ R\left({d\theta\over dt}\right)\right]^2 +
    \left({dR\over dt}\right)^2\right\}^{1/2}~~.
  \label{eq:vel}
\end{equation}
The time derivative of the velocity vector provides the acceleration
vector, ${\bf a}=(d{\bf v}/dt)$.

Remarkably, in the special case where the radial speed is negligible,
$\vert dR/dt\vert \ll R\vert d\theta/dt\vert$, I find that the lower
limit on the object's mass is independent of distance,
\begin{equation}
M > {4\pi f \over \vert ({d\theta/dt})\times(d^2\theta/dt^2)\vert }~~.
\label{eq:nodis}
\end{equation}

The physical size of the object can be derived as the product of its
angular size times its distance, $(\Delta \theta)R$. The minimum mass
could then be used to derive the minimum mass per unit volume, or mass
density $\rho$, of the object.

\section{Discussion}

If the measured velocity and acceleration of a technological object
are outside the flight characteristics and performance envelopes of
drones or airplanes, then the object would be classified by the {\it
  Galileo Project}'s research team as an outlier. In such a case, it
would be interesting to calculate the minimum mass density of the
object. If the result exceeds normal solid densities, then the object
would qualify as anomalous, a UAP. Infrared emission by the object
would be a source of confusion, unless the object is resolved and the
emission from it can be separated from the heated air around it.

All flying objects made by humans have a volume-averaged mass density
$\langle\rho\rangle$ which is orders of magnitude below $22.6~{\rm
  g~cm^{-3}}$, the density of Osmium - which is the densest metal
known on Earth. A UFO with a higher mass density than Osmium would
have to carry exotic material, not found on Earth.

By summer 2025, there will be three {\it Galileo Project}
observatories operating in three different states within the U.S. and
collecting data on a few million objects per year. With new
quantitative data on infrared luminosities, velocities and
accelerations of technological objects, it would be possible to check
whether there are any UFOs denser than Osmium.

\bigskip
\bigskip
\bigskip
\bigskip
\section*{Acknowledgements}

This work was supported in part by the {\it Galileo Project} at
Harvard University.
 
\bigskip
\bigskip
\bigskip

\bibliographystyle{aasjournal}
\bibliography{t}
\label{lastpage}
\end{document}